\documentclass{aa}
\usepackage{graphicx,txfonts,natbib}
\usepackage[switch]{lineno}
\usepackage{hyperref}

\def \kms {{\rm km\;s$^{-1}$}}

\def \arcsec {$^{''}$}
\def \ha {H$\alpha$}
\def \lya {Ly$\alpha$}

\def \tb {$T_{b}$}

\graphicspath{{./}}

\begin{document}


\title{Radio dimming associated with filament eruptions in the meter and decimeter wavebands}

\author{Zhenyong Hou\inst{1}
          \and
          Hui Tian\inst{1,2}
          \and
          Jingye Yan\inst{2,3,4}
          \and
          Maria S. Madjarska\inst{5,6,7}
          \and
          Jiale Zhang\inst{1}
          \and
          Yu Xu\inst{1}
           \and
          Hechao Chen\inst{8}
          \and
          Zhao Wu\inst{9}
          \and
          Lin Wu\inst{2,3,4}
          \and
          Xuning Lv\inst{2,3,4}
          \and
          Yang Yang\inst{2,3,4}
          \and
          Yujie Liu\inst{2,3,4}
          \and
          Li Deng\inst{2,3,4}
          \and
          Li Feng\inst{10}
          \and
          Ye Qiu\inst{11}
}

\institute{School of Earth and Space Sciences, Peking University, Beijing, 100871, China\\ \email{huitian@pku.edu.cn}
         \and
         Key Laboratory of Space Weather, Chinese Academy of Sciences, Beijing, 100871, China\\ \email{yanjingye@nssc.ac.cn}
         \and
         National Space Science Center, Chinese Academy of Sciences, Beijing, 100871, China
         \and
         Radio science and technology center ($\pi$ center), Chengdu, 610041, China
         \and
         Korea Astronomy and Space Science Institute, 34055, Daejeon, Republic of Korea
         \and 
        Max Planck Institute for Solar System Research, Justus-von-Liebig-Weg 3, 37077, G\"ottingen, Germany
         \and
         Space Research and Technology Institute, Bulgarian Academy of Sciences, Acad. Georgy Bonchev Str., Bl. 1, 1113, Sofia, Bulgaria
         \and 
         School of Physics and Astronomy, Yunnan University, Kunming, 650050, China
         \and 
         School of Space Science and technology, Shandong University, Weihai, 264209, China
         \and
         Key Laboratory of Dark Matter and Space Astronomy, Purple Mountain Observatory, CAS, Nanjing, 210023, China
         \and
         School of Astronomy and Space Science, Nanjing University, Nanjing, 210023, China
}

\date{Received date, accepted date}

\abstract
{Filament eruptions are considered to be a common phenomenon on the Sun and other stars, yet they are rarely directly imaged in the meter and decimeter wavebands.
Using imaging data from the DAocheng solar Radio Telescope (DART) in the 150--450 MHz frequency range, we present two eruptive filaments that manifest as radio dimmings (i.e., emission depressions).
Simultaneously, portions of these eruptive filaments are discernible as dark features in the chromospheric images.
The sun-as-a-star flux curves of brightness temperature, derived from the DART images, exhibit obvious radio dimmings.
The dimming depths range from 1.5\% to 8\% of the background level and show a negative correlation with radio frequencies and a positive correlation with filament areas.
Our investigation suggests that radio dimming is caused by free-free absorption during filament eruptions obscuring the solar corona.
This may provide a new method for detecting stellar filament eruptions.}

\authorrunning{Hou et al.}
\titlerunning{Radio dimming associated with filament eruptions}

\maketitle

\section{Introduction}
\label{sec:intro}

Solar filament eruptions have been extensively studied through observations in the extreme-ultraviolet (EUV) and \ha\ wavebands \citep[e.g.,][]{2007SoPh..240...77J,2007ApJ...668..533N,2008ApJ...681L..37L,2011ApJ...739...43L,2011MNRAS.414.2803Y,2012ApJ...760...81K,2022ApJ...931...76X,2023ApJ...953...68L,2023ApJ...957...58W}, as well as through magnetohydrodynamic simulations \citep[e.g.,][]{2005ApJ...630L..97T,2022ApJ...940..119Y,2023ApJS..266....3G,2023ApJ...958...25G}.
When a filament successfully erupts from the Sun, it evolves into a coronal mass ejection \citep[CME, e.g.,][]{2006ApJ...651.1238Z,2008ApJ...674..586S,2009SoPh..259...13G,2020RAA....20..166C}, which is the primary driver of solar storms that significantly affect the space environment throughout the Solar System.
Conversely, if a filament eruption fails due to an interaction with overlying magnetic fields, it returns to the solar surface without producing a CME \citep[e.g.,][]{2003ApJ...595L.135J,2009ApJ...696L..70L,2014ApJ...784L..36Y,2022ApJ...933L..38D}.
Filament eruptions vary in scale, ranging from a few to thousands of megameters \citep[e.g.,][]{2020ApJ...902....8C,2022A&A...660A..45M,2024ApJ...961L..30Q,2024ApJ...974..205S,2024SoPh..299...85S,2024ApJ...974..123W,2024arXiv241001123S}, and they are often associated with other phenomena such as EUV waves \citep[e.g.,][]{2020ApJ...894..139Z,2022SoPh..297..153D,2024ApJ...968...85Z} and type II radio bursts \citep[e.g.,][]{2014SoPh..289.2123K,2022ApJ...926L..39D}.
Magnetic reconnection plays a critical role in these eruptions \citep[e.g.,][]{2000JGR...105.2375L,2016NatCo...711837X,2023ApJ...959...69H,2024A&A...684A..14W,2024A&A...687A.190H}.
Similar to the Sun, filament eruptions have also been found on other stars \citep[e.g.,][]{2022ApJ...933...92C,2023ScSnT..53.2021T,2024ApJ...961...23N,2022A&A...663A.140L,2025ApJ...978L..32L}.
However, observational evidence for stellar filament eruptions and CMEs remains limited.

Solar filaments have also been observed in the millimeter and centimeter wavebands.
In these wavebands, filaments exhibit a similar shape and position to those in \ha\ images but appear broader and longer \citep[e.g.,][]{1994PASJ...46..205H,1996NewA....1..207G}.
Both quiescent and eruptive filaments manifest as emission depressions on the solar disk, while they appear as bright features above the solar optical limb \citep[e.g.,][]{1998ApJ...498L.179G,1999ApJ...510..466H,2013PASJ...65S..11G}.
Studies have shown that eruptive filaments have an emission enhancement after the beginning of the eruptions in the images of the millimeter and centimeter wavebands \citep{1999ApJ...510..466H,2013PASJ...65S..11G}.
\cite{2008SoPh..253..263G} conducted a sun-as-a-star analysis of an eruptive filament and found radio dimming associated with the eruption in the 1--5 GHz frequency range.
The dimming duration and depth increased toward lower frequencies.
However, to our knowledge, in the meter and decimeter wavebands, only \cite{2001A&A...374..316M} and \cite{2002A&A...387..317M} have reported two events related to filament eruptions using imaging data taken by the Nan\c{c}ay Radioheliograph \citep[NRH,][]{1997LNP...483..192K}, and they did not perform a detailed analysis of the emission depression in the radio images.

Using imaging data in the meter and decimeter wavebands from the DAocheng solar Radio Telescope \citep[DART,][]{2023NatAs...7..750Y}, we investigated two events of obscuration dimmings caused by filament eruptions.
The sun-as-a-star flux curves in the frequency range of 150--450 MHz show a significant depression of brightness temperature (\tb) below the pre-eruption levels during the filament eruptions. This indicates that radio dimming in the meter and decimeter wavebands could be a useful tool for detecting stellar filament eruptions.
We describe the observations in Section\,\ref{sec:obs}, explain the methodology of the analysis in Section\,\ref{sec:met}, present the analysis results in Section\,\ref{sec:res}, discuss the results in Section\,\ref{sec:dis}, and summarize our findings in Section\,\ref{sec:sum}.

\section{Observations}
\label{sec:obs}

\begin{figure*}
\centering
\includegraphics[trim=0.0cm 0.3cm 0.0cm 0.0cm,width=0.9\textwidth]{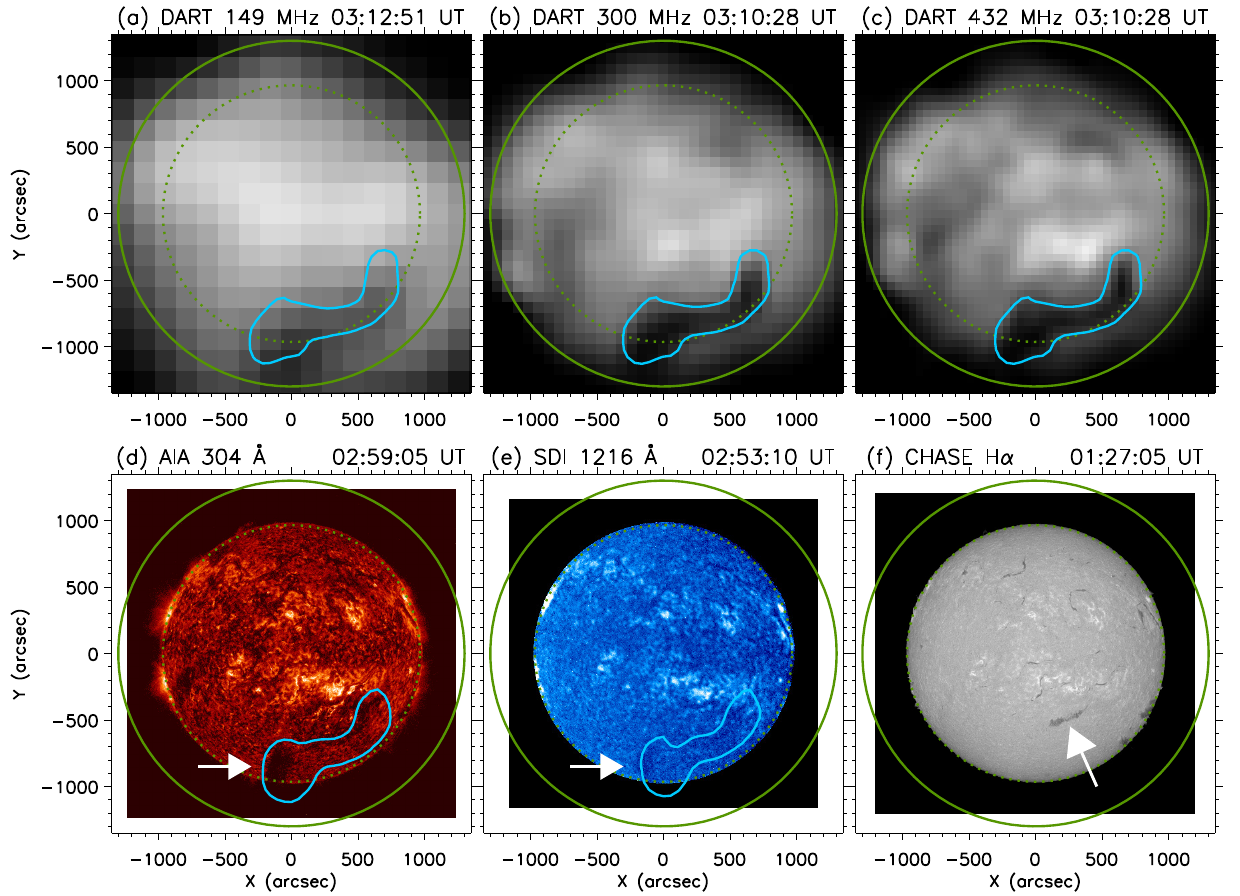}
\caption{Overview of the filament eruption on 2024 March 17.
(a)--(c) DART images at 149 MHz, 300 MHz, and 432 MHz, respectively.
(d)--(f) Images of the AIA 304 \AA, SDI \lya, and CHASE \ha\ line-center wavebands, respectively.
The solid green lines outline the integration circle with a radius of 1300\arcsec\ for obtaining the radio flux curves, while the dotted green lines indicate the solar optical limb in the \lya\ waveband.
The cyan contours in (a)--(e) outline the eruptive filament in the 432 MHz images.
The white arrows in (d) and (e) indicate the dark regions associated with the eruptive filament, while the white arrow in (f) marks the quiescent filament before the eruption.
An online animation illustrates the eruptive filament over a duration of $\sim$2 hours, from 02:01~UT to~03:58 UT.}
\label{fig:overview0317}
\end{figure*}

\begin{figure*}
\centering
\includegraphics[trim=0.0cm 0.3cm 0.0cm 0.0cm,width=0.9\textwidth]{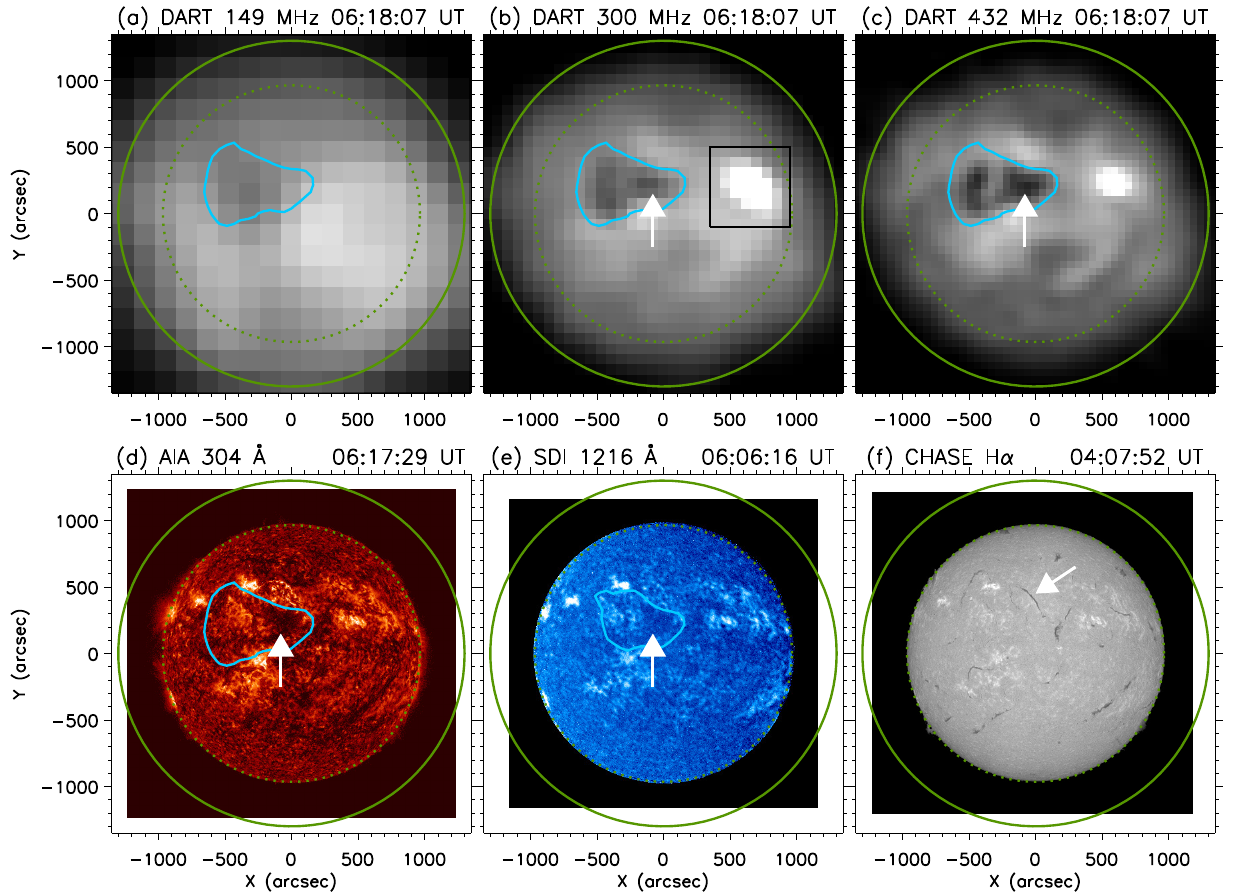}
\caption{Similar to Figure\,\ref{fig:overview0317} but for the filament eruption on 2024 April 11.
The box in (b) outlines a bright region covering several patches of plage areas.
An online animation illustrates the eruptive filament over a duration of $\sim$3.5 hours, from 04:00 UT to 07:30 UT.}
\label{fig:overview0411}
\end{figure*}

The data used in this study were taken by DART, an instrument of the Meridian Project of China \citep{2024SpWea..2203972W} designed to detect solar radio emission from the high corona through imaging and spectral observations.
DART detected two large filament eruptions originating from quiet Sun regions on 2024 March 17 and 2024 April 11, respectively.
We analyzed DART imaging data in 16 wavebands (i.e., 149, 164, 190, 205, 223, 238, 285, 300, 309, 324, 366, 381, 399, 414, 432, and 447~MHz) within the frequency range of 150--450~MHz.
The time cadences of the imaging observations for the events on 2024 March 17 and 2024 April 11 are 10 s and 3 minutes, respectively. 
The angular resolution of the radio images changes from $\sim$100\arcsec\ at 447 MHz to $\sim$300\arcsec\ at 149 MHz.

We also analyzed simultaneous observations from other instruments, including the Atmospheric Imaging Assembly \citep[AIA,][]{2012SoPh..275...17L} on board the Solar Dynamics Observatory \citep[SDO,][]{2012SoPh..275....3P}, the Solar Disk
Imager (SDI) of the Lyman alpha (\lya) Solar Telescope \citep[LST,][]{2019RAA....19..158L,2024SoPh..299..118C} on board the Advanced Space-based Solar Observatory \citep[ASO-S,][]{2023SoPh..298...68G} mission, and the \ha\ Imaging Spectrograph \citep[HIS,][]{2022SCPMA..6589603Q} on board the Chinese \ha\ Solar Explorer \citep[CHASE,][]{2022SCPMA..6589602L}.
The AIA 304\,\AA\ images used to examine the eruptive filaments have a 12 s cadence and a 0.6\arcsec\ pixel size.
The SDI \lya\ images that also recorded the eruptive filament have a time cadence of 3 minutes and a pixel size of 0.5\arcsec.
We also analyzed  CHASE \ha\ line-center images with a time cadence of 1--2 minutes and a pixel size of 1.0\arcsec to identify the quiescent filaments prior to their eruptions.

\section{Methodology}
\label{sec:met}

\begin{figure*}
\centering
\includegraphics[trim=0.0cm 0.3cm 0.0cm 0.0cm,width=0.7\textwidth]{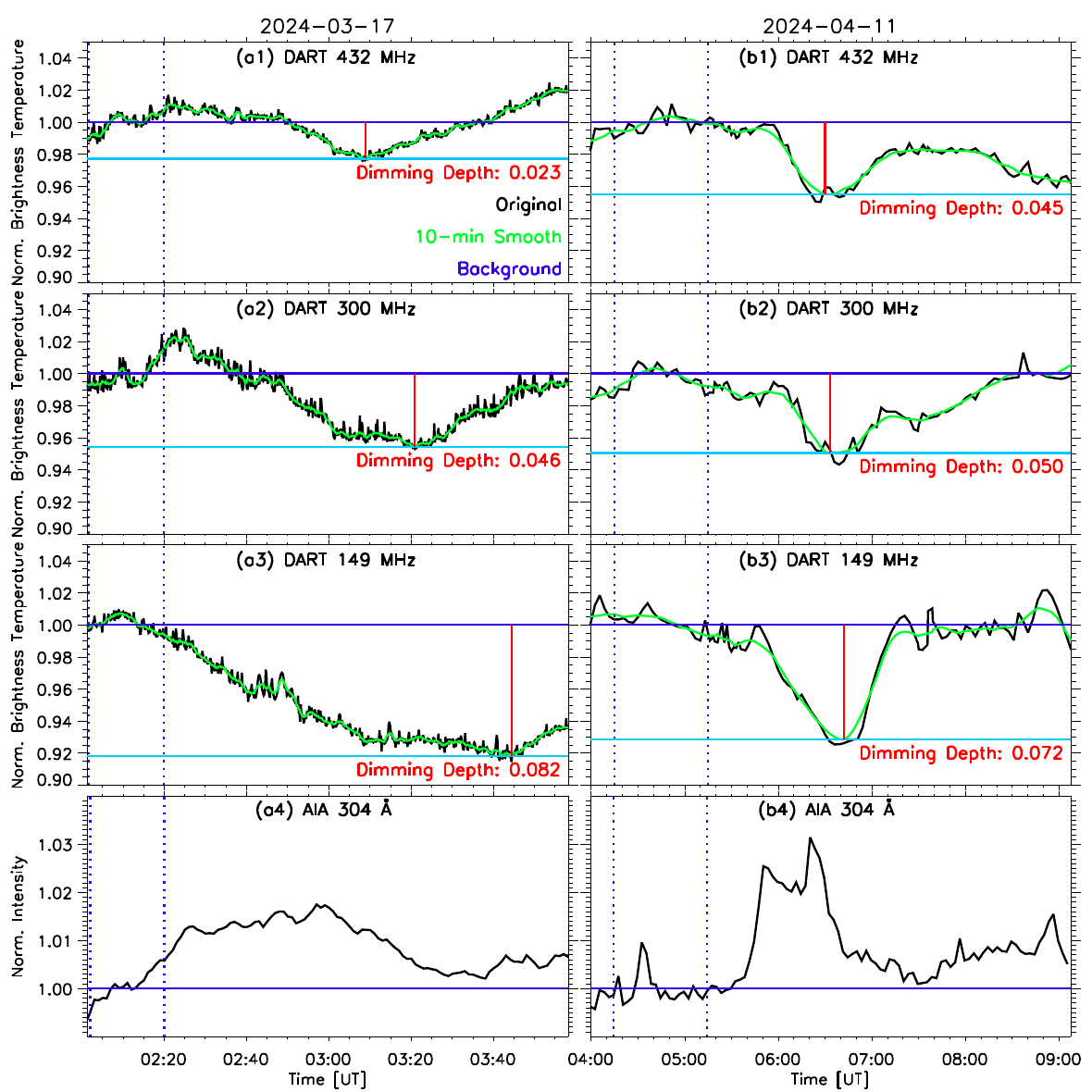}
\caption{Temporal evolution of the normalized sun-as-a-star fluxes.
The black lines represent the original flux curves, while the green lines show the flux curves smoothed over a 10-minute interval.
The vertical blue lines indicate the pre-eruption period, with the horizontal blue lines marking the pre-eruption levels.
The cyan lines denote the minimum flux values, and the vertical red lines highlight the dimming depths.
}
\label{fig:lc}
\end{figure*}

For this study, we analyzed two filament eruptions and the associated radio dimmings using the DART imaging data at 150--450 MHz.
To conduct a detailed investigation, we performed a sun-as-a-star flux analysis across all wavebands.

\subsection{Sun-as-a-star emission flux}
\label{subsec:tblc}

Using radio images at a specific frequency, we calculated the sun-as-a-star flux curve ($F_{total(T_{b})}$) by integrating the \tb\ values of all pixels within a circular region centered on the solar disk with a radius of 1300\arcsec\ (marked by the solid green lines in Figures\,\ref{fig:overview0317} and \ref{fig:overview0411}) at each observation time.
To compare with a background, we defined the pre-eruption period as the time preceding the eruption. We used the average value from this period as the pre-eruption level of radio emission ($F_{total(T_{b_{0}})}$).

Numerous spikes appear in the sun-as-a-star flux curves ($F_{total(T_{b})}$).
They are caused by the plage emission (see the bright region marked by the box in Figure\,\ref{fig:overview0411}) or other solar activities.
To remove these spikes, we first applied a smoothing window of 30 minutes to the flux curves ($F_{total(T_{b}, smooth30)}$) and calculated the relative deviations between the smoothed and original flux curves through $(F_{total(T_{b}, smooth30)} - F_{total(T_{b})})/F_{total(T_{b_{0}})}$.
Then, we excluded the observational times when the absolute values of the relative deviations exceed 0.01.

We then applied a smoothing window of 10 minutes to the despiked radio flux curves ($F_{total(T_{b}, smooth10)}$) to mitigate fluctuations in the flux curves.
After dividing $F_{total(T_{b}, smooth10)}$ by $F_{total(T_{b_{0}})}$, we derived the temporal variations of the normalized total fluxes ($F_{norm(T_{b})}$, see the green lines in Figure\,\ref{fig:lc}).
For further analysis, we defined the dimming depth (see the vertical red lines in Figure\,\ref{fig:lc}) as the percentage decrease of the flux depression relative to the pre-eruption level by 1 - $F_{norm(T_{b})}$.
Using the same method, we also calculated the temporal variations of the normalized AIA 304 \AA\ intensity, but with an integration circle of 1200\arcsec radius.

\subsection{Identification of the eruptive filaments}
\label{subsec:fila_area}

For the subsequent analysis, we also examined the relationship between the dimming depth and the filament area.
The DART 432 MHz images were used to measure the filament areas, as this waveband clearly captured the eruptive filaments.
For each event, we enhanced the signal-to-noise ratio by applying a temporal smoothing window of 10 minutes to the 432 MHz radio images.
A series of base-difference images were then produced by subtracting the background image from the smoothed images.
To identify the filament regions, we applied a threshold value to the base-difference images.
Examples of these filament regions are shown in Figures\,\ref{fig:overview0317} and \ref{fig:overview0411}, marked by cyan contours.
These contours outline the filament regions well across all frequencies, confirming that the filament pixels identified at 432 MHz can be used to measure the filament areas.

\section{Results}
\label{sec:res}

\begin{figure*}
\centering
\includegraphics[trim=0.0cm 0.0cm 0.0cm 0.0cm,width=0.8\textwidth]{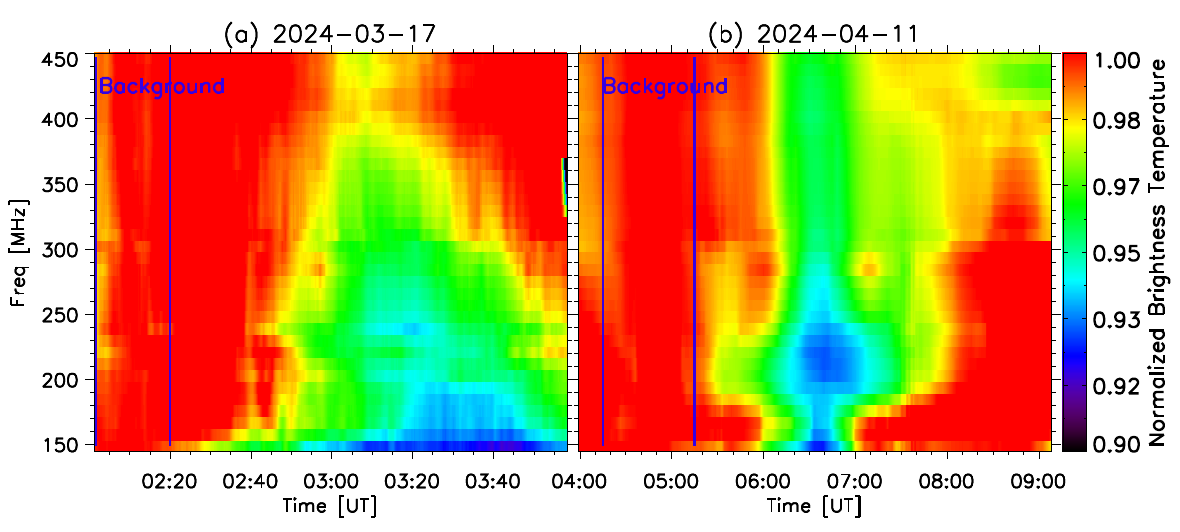}
\caption{DART radio dynamic spectra combining the normalized sun-as-a-star flux curves in the meter and decimeter wavebands.
The blue lines mark the pre-eruption period.
The flux curve at each frequency has been normalized to the corresponding pre-eruption level.}
\label{fig:sp}
\end{figure*}

Figures\,\ref{fig:overview0317} and \ref{fig:overview0411} provide an overview of these two filament eruptions.
The eruptive filaments are clearly visible in the DART images, particularly at higher frequencies, where they appear as flux depressions both on the solar disk and slightly above the solar optical limb (also see the associated animations of Figures\,\ref{fig:overview0317} and \ref{fig:overview0411}).
The DART radio images at higher frequencies can reveal fine structures of the eruptive filament.
For example, on 2024 April 11, two different components of the eruptive filament appear successively in the radio images at higher frequencies as seen in Figures\,\ref{fig:overview0411}(b) and \ref{fig:overview0411}(c) (also see the associated animations).
Falling plasma following the second component near the foot of the eruptive filament is also detected in the radio images.
These two events are not associated with obvious flares.
Some dark structures related to the eruptive filaments, indicated by the white arrows in Figures\,\ref{fig:overview0317}(d)--(e) and \ref{fig:overview0411}(d)--(e), are also detected in the AIA 304\,\AA\ and the SDI \lya\ wavebands.
This suggests that the eruptive filaments primarily consist of cool plasma that may not undergo heating during the eruption.
The CHASE \ha\ line-center images reveal the pre-eruption filaments located in quiet Sun regions (see Figures\,\ref{fig:overview0317}(h) and \ref{fig:overview0411}(h)).
The lengths of these quiescent filaments, as measured from the \ha\ line-center images, are 210--230 Mm.
To identify the associated CMEs, we checked the coronagraph images from the JHelioviewer \citep{2017A&A...606A..10M} taken by the Large Angle Spectroscopic Coronagraph
  \citep[LASCO,][]{1995SoPh..162..357B} on board the Solar and Heliospheric Observatory \citep[SOHO,][]{1998GeoRL..25.2465T}, and confirmed that both events are associated with a CME.

\begin{figure*}
\centering
\includegraphics[trim=0.0cm 0.3cm 0.0cm 0.0cm,width=1.0\textwidth]{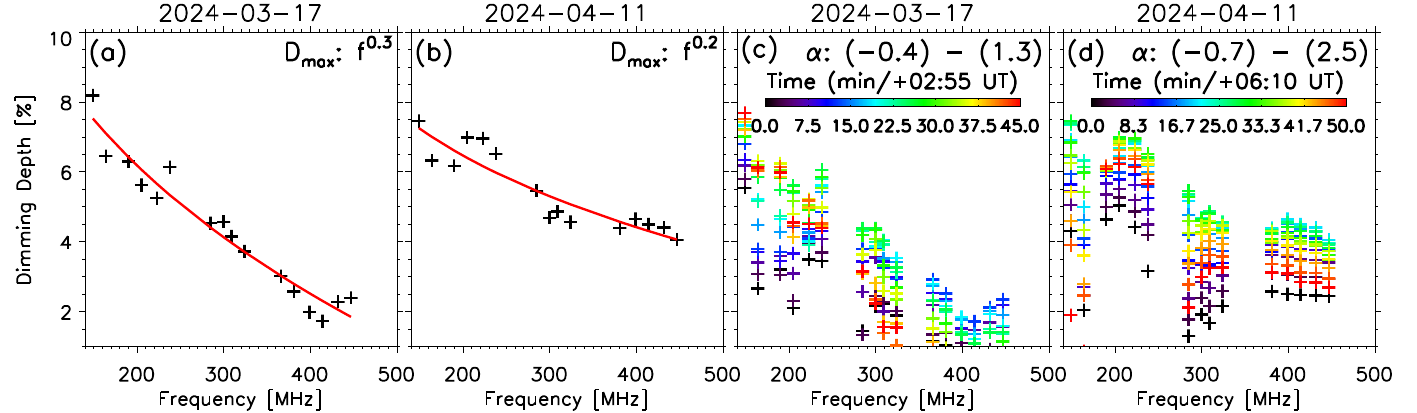}
\caption{Scatter plots illustrating the relationship between radio frequency and dimming depth.
(a)--(b) Scatter plots for $D_{max}$.
The red lines indicate the linear fits between $f^{\alpha}$ and $D_{max}$.
(c)--(d) Scatter plots for dimming depth over time.}
\label{fig:depth_freq}
\end{figure*}

\subsection{Sun-as-a-star dimming}
\label{subsec:lc}

The DART images and movies reveal radio dimmings caused by the eruptive filaments in the meter and decimeter wavebands.
These radio dimmings are also evident in the sun-as-a-star flux curves, as explained in Section\,\ref{subsec:tblc} and shown in Figure\,\ref{fig:lc}, where the normalized \tb\ values just after the eruptions fall below the pre-eruption levels.
For the event on 2024 March 17 (2024 April 11), compared to the pre-eruption levels, the dimming depths at 432 MHz, 300 MHz, and 149 MHz were measured as 0.023 (0.045), 0.046 (0.05), and 0.082 (0.072), respectively.
The radio dimmings have a larger dimming depth and last longer at lower frequencies.
This trend is further illustrated in Figure\,\ref{fig:sp}, which shows the normalized radio dynamic spectra as a combination of the normalized sun-as-a-star flux curves across all frequencies.

As mentioned in Section\,\ref{subsec:tblc}, we also examined the sun-as-a-star flux curves in the AIA 304 \AA\ waveband, shown in the bottom row of Figure\,\ref{fig:lc}.
However, despite the presence of dark structures associated with the eruptive filaments in this waveband, the emission levels after the eruptions remain above the pre-eruption values.
This could be attributed to the relatively weak obscuration of bright active regions by the eruptive filaments.

\begin{figure*}
\centering
\includegraphics[trim=0.0cm 0.3cm 0.0cm 0.0cm,width=0.80\textwidth]{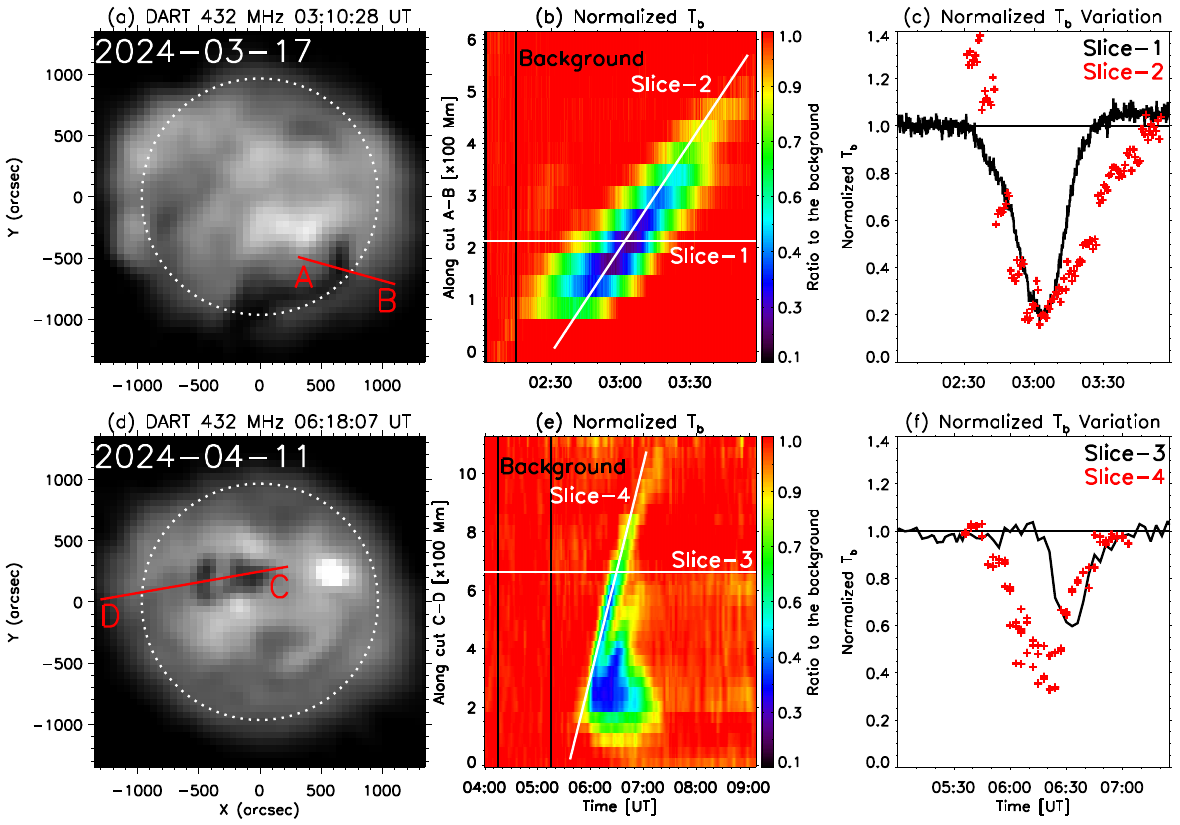}
\caption{Evolution of the normalized flux at 432 MHz.
Left: DART 432 MHz images of the two eruptions.
Middle: The time-distance diagrams along the cuts are shown in the left column.
Right: Normalized flux variations along the slices indicated in the middle column.}
\label{fig:Tvariation}
\end{figure*}

\begin{figure*}
\centering
\includegraphics[trim=0.0cm 0.3cm 0.0cm 0.0cm,width=1.0\textwidth]{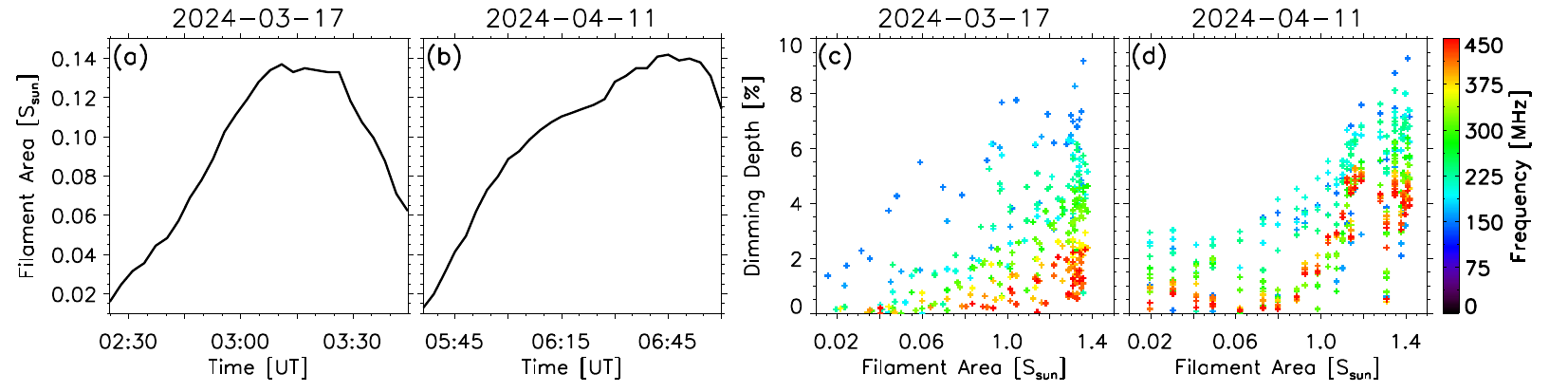}
\caption{Relationship between the filament area and dimming depth.
(a)--(b) Temporal variations of the filament areas at 432 MHz.
(c)--(d) Scatter plots between filament areas and dimming depths at different frequencies represented by different colors.
The color bar illustrates the frequency variations.}
\label{fig:depth_area}
\end{figure*}

\subsection{Relationship between the dimming depth and the radio frequency}
\label{subsec:depth_freq}

Here, we analyze the relationship between dimming depths and radio frequencies, as shown in Figure\,\ref{fig:depth_freq}.
Figures\,\ref{fig:depth_freq}(a) and \ref{fig:depth_freq}(b) display scatter plots of the maximum dimming depths ($D_{max}$) against the frequencies.
We note that $D_{max}$ refers to the maximum dimming depth observed during the dimming period at each frequency and it appears to exhibit a negative correlation with frequency in the range of 150--450 MHz.
For the event on 2024 March 17, the DART imaging data with a 3-minute time cadence were utilized.
A power-law function, expressed as $D_{max} = a\cdot f^{\alpha}+b$, where f is the frequency in units of hertz, was applied to model the relationship between the maximum dimming depths and the frequencies.
The correlation coefficient between $D_{max}$ and $f^{\alpha}$ was maximized when the power-law index $\alpha$ was 0.3 for the 2024 March 17 event and 0.2 for the 2024 April 11 event.

Figures\,\ref{fig:depth_freq}(c) and \ref{fig:depth_freq}(d) illustrate the relationship between dimming depths and frequencies over time.
The relationships for both events are similar, showing a negative correlation between dimming depths and frequencies.
The power-law index $\alpha$ varies from -0.4 to 1.3 for the 2024 March 17 event and from $-$0.7 to 2.2 for the 2024 April 11 event.

\subsection{Evolution of the brightness temperature}
\label{subsec:evo_tb}

To examine the \tb\ evolution in the eruptive filaments, we placed two cuts along the propagation paths of the filaments and generated the corresponding time-distance diagrams from the DART images at 432 MHz.
The results are presented in Figure\,\ref{fig:Tvariation}.

For the filament eruption on 2024 March 17, Cut A--B was placed along the propagation path of its western part.
Figure\,\ref{fig:Tvariation}(b) presents the resulting time-distance diagram, where the \tb\ value at each position was normalized to the pre-eruption level.
The propagation of the eruptive filament is clearly visible in the time-distance diagram, with a velocity of $\sim$110 \kms.
Two slices were selected (indicated by the white lines in Figure\,\ref{fig:Tvariation}(b)), and the variations of the normalized \tb\ along these slices are displayed in Figure\,\ref{fig:Tvariation}(c).
The black line shows the \tb\ variation along slice-1 for a fixed pixel on the solar disk, revealing a rapid on-off-on pattern.
At this fixed pixel, the \tb\ remains stable before the eruption, corresponding to the pre-eruption level.
During the eruption, the eruptive filament moves quickly across this pixel, leading to a rapid \tb\ decrease and recovery.
The \tb\ variation along slice-2 reveals a pattern similar to the sun-as-a-star flux curve.
During the eruption, the \tb\ rapidly decreases to well below the pre-eruption level and gradually returns to the pre-eruption level.
Figure\,\ref{fig:Tvariation}(d)--(f) presents analogous results for the eruption on 2024 April 11.
The \tb\ evolution in the eruptive filaments at other frequencies exhibits a similar behavior.

The upward ejecting and falling components of the eruptive filament on 2024 April 11 also appear in the time-distance diagram as shown in Figure\,\ref{fig:Tvariation}(e).
These components were also captured by the AIA 304\,\AA\ images, but not as clear as in the radio images.

\subsection{Relationship between the dimming depth and the filament area}
\label{subsec:depth_area}

We also analyzed the relationship between the dimming depths and filament areas for both events.
First, we determined the filament areas during the eruptions using the method described in Section\,\ref{subsec:fila_area}.
For the 2024 March 17 event, data with a 3-minute cadence were utilized.
The measured filament areas for both events are shown in Figures\,\ref{fig:depth_area}(a) and \ref{fig:depth_area}(b), where the filament area can reach up to 13\% of the solar disk area.
Figures\,\ref{fig:depth_area}(c) and \ref{fig:depth_area}(d) present scatter plots illustrating the relationship between filament areas and dimming depths at different frequencies.
The results indicate a positive correlation between dimming depths and filament areas for both events.

For the 2024 March 17 event, the filament areas exceeded 13\% of the solar disk area between 03:05 and 03:27 UT.
This period corresponds to the maximum dimming depths at higher frequencies (350--450 MHz), but it occurs before the maximum dimming depths at lower frequencies (150--200 MHz), as shown in Figure\,\ref{fig:sp}(a).
This suggests that the changes in the filament area may not be the sole factor influencing the variations of the dimming depths.
For example, as mentioned in Section\,\ref{subsec:tblc}, bright regions (caused by plage regions or solar activities) within the integration circle in the radio images affects the measurements of the dimming depths. 
In addition, Figures\,\ref{fig:depth_area}(c) and \ref{fig:depth_area}(d) demonstrate that radio dimmings of these two events are observable in the sun-as-a-star flux curves when the expanding filaments occupy more than $\sim$2\% of the solar disk area.


\begin{figure*}
\centering
\includegraphics[trim=0.0cm 1.5cm 0.0cm 0.0cm,width=0.8\textwidth]{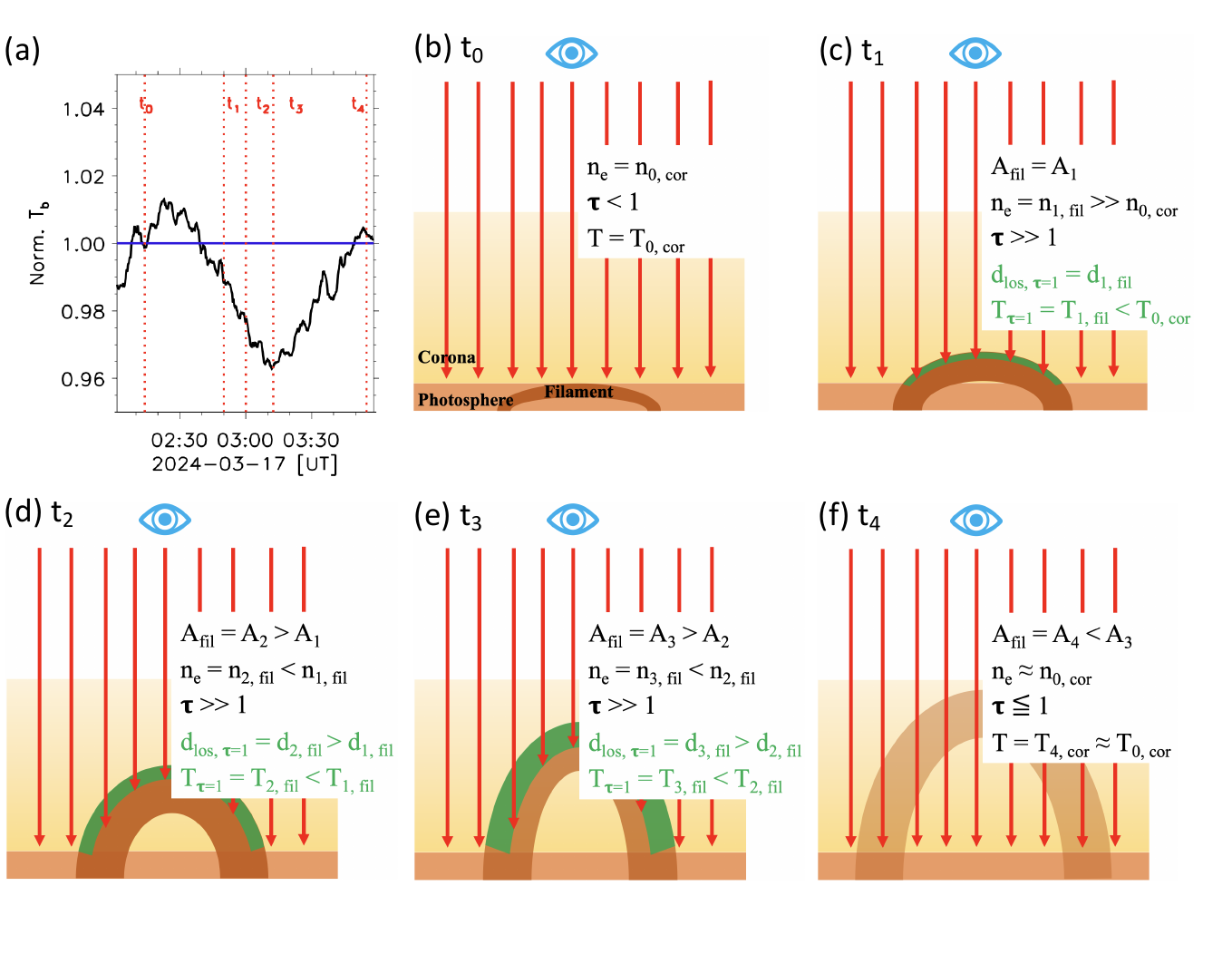}
\caption{Cartoon illustrating a possible connection between a radio dimming and filament eruption.
(a) Evolution of the normalized flux at 324 MHz on 2024 March 17, representing both the sun-as-a-star flux curves in Figure\,\ref{fig:lc} and the flux variations in the right column of Figure\,\ref{fig:Tvariation}.
In (a), the blue line marks the pre-eruption level.
(b)--(f) The filament eruption process and the changes of the $d_{los, \tau=1}$ (indicated by the red arrows) and the obscuration area (A).
The times t$_{0}$--t$_{4}$ correspond to the times shown in (a).
In (b)--(f), "cor" and "fil" represent "corona" and "filament," respectively.}
\label{fig:cartoon}
\end{figure*}

\section{Discussion}
\label{sec:dis}

In non-flaring regions of the solar corona, radio emission is primarily generated by free-free emission in the meter and decimeter wavebands.
The optical depth $\tau$ for thermal electrons can be expressed as \citep[e.g.,][]{1985ARA&A..23..169D,2020FrASS...7...57N}
\[\tau = \kappa \times d_{los} \approx\ 0.2 \times \frac{n_{e}^{2}}{f^{2}T_{e}^{3/2}} \times d_{los}, \]
where $\kappa$ is the free-free absorption coefficient, $d_{los}$ is the line of sight (LOS)  thickness of the source, $n_{e}$ is the electron number density, $f$ is the frequency in units of hertz, and $T_{e}$ is the electron temperature.
Here, we present two eruptive filaments observed in the DART imaging data in the meter and decimeter wavebands, covering a frequency range of 150--450 MHz.
These wavebands image the solar corona.
For one eruptive filament with $n_{e} \approx 10^{10}$ cm$^{-3}$, $T_{e} \approx$ 8000 K, and $d_{los} \approx$ 50 Mm observed at 432 MHz, we got $\tau = 7.5 \times 10^{5} \gg 1$, indicating that the filament is optically thick.
In contrast, considering one coronal structure with $n_{e} \approx 10^{8}$ cm$^{-3}$, $T_{e} \approx$ 10$^{6}$ K, and $d_{los} \approx$ 50 Mm observed at 432 MHz, the coronal structure was found to be optically thin, as its $\tau$ was measured to be 0.054 $\ll$ 1.
These results suggest that as the filament expands into the corona, and that it obscures and absorbs radio emission from the underlying corona.
Meanwhile, the eruptive filament, composed of cooler and denser plasma, emits weakly, resulting in a flux depression in radio images across the 150--450 MHz frequency range.
This process is referred to as free-free absorption.
When the expanding filament occupies more than 2\% of the solar disk area, radio dimming becomes detectable in the sun-as-a-star fluxes.

The optical depth is inversely proportional to the square of frequency, that is, $\tau \sim f^{-2}$.
For the lower frequency, it corresponds to a larger optical depth, suggesting that the imaged filament can be more efficient to obscure background emission at lower frequencies.
This indicates that the variation of the optical depth across different frequencies may, to some extent, explain the negative relationship between the dimming depth and the frequency.

In this study, we present significant radio dimmings observed in both the imaging data and the sun-as-a-star flux curves in the 150--450 MHz frequency range.
This is the first demonstration of radio dimming using the sun-as-a-star analysis in the meter and decimeter wavebands.
\cite{2001A&A...374..316M} and \cite{2002A&A...387..317M} previously reported two eruptive filaments observed in this frequency range using NRH imaging data.
However, they did not employ a sun-as-a-star analysis or examine the properties of the emission depression caused by the eruptive filaments.

Previous studies have reported quiescent and eruptive filaments and prominences in the millimeter and centimeter wavebands.
These filaments typically appear as emission depressions on the solar disk but as bright features above the solar optical limb \citep[e.g.,][]{1998ApJ...498L.179G,1999ApJ...510..466H,2013PASJ...65S..11G}.
While eruptive filaments have been observed to be heated during the eruptions, exhibiting an emission value higher than that of quiescent filaments \citep{1999ApJ...510..466H,2013PASJ...65S..11G}, the eruptive filaments in our study display different characteristics.
Specifically, they show no signs of heating and appear as depressions above the solar optical limb.
\cite{2008SoPh..253..263G} conducted a sun-as-a-star analysis of an eruptive filament, identifying radio dimming in the 1--5 GHz frequency range.
Similar to our findings, the duration and depth of the dimming increased toward lower frequencies.

Coronal dimmings are typically observed as reduced emission in the EUV and soft X-ray wavebands \citep[e.g.,][]{1997ApJ...491L..55S,2003A&A...397.1057Z,2007ApJ...660.1653M,2018ApJ...855..137D,2020A&A...633A.142Z,2023A&A...678A.166C,2023ApJ...952L..29W}, and they are often interpreted as mass depletion caused by escaping CMEs \citep[e.g.,][]{1996ASPC..111..379H,2009ApJ...702...27J,2012ApJ...748..106T}.
Their chromospheric counterparts, such as \ha\ dimmings, have also been reported \citep{2003ApJ...597L.161J,2007SoPh..240...77J}.
However, the radio dimmings observed in our study arise from a different mechanism.
In this case, the eruptive filaments propagate across the solar disk and absorb background emission, a process known as obscuration dimming \citep{2014ApJ...789...61M}.
This mechanism can occasionally produce dimming in sun-as-a-star fluxes in the AIA~304 \AA\ waveband \citep{2024ApJ...970...60X}.

In our study, we found a positive correlation between the dimming depths and the filament areas.
An additional factor affecting the variations of the dimming depths is opacity.
This is because the absorption depth we observe is likely a trade-off between the filament's expansion and its transparency changes.
As the filament erupts, it expands and covers a larger region, causing stronger background absorption.
On the other hand, the filament also grows less dense, changing its optical depth from $\tau \gg$ 1 to $\tau \sim$ 1 or $<$ 1 and reducing absorption.
Figure\,\ref{fig:cartoon} illustrates the radio dimming linked to changes in the obscuration area (A), optical depth $\tau$, and LOS depth $d_{los}$ during the filament eruptions.
For example, the sun-as-a-star flux curve ($F_{norm(T_{b})}$) at 324 MHz on 2024 March 17 is shown in Figure\,\ref{fig:cartoon}(a), with five time markers ($t_{0}-t_{4}$) corresponding to the changes of A, $\tau$, and $d_{los}$ depicted in Figure\,\ref{fig:cartoon}(b)--(f).
The flux enhancement after $t_{0}$ is caused by the flare associated with the eruptive filament.
At $t_{0}$ in Figure\,\ref{fig:cartoon}(b), prior to the eruption, $\tau$ is much less than one, indicating that the region is optically thin and allows for the capture of coronal emission ($T_{0}$).
At the onset of the eruption ($t_{1}$ in Figure\,\ref{fig:cartoon}(c)), the eruptive filament expands into the corona with the obscuration area of A$_{1}$.
The filament density is high, and $\tau$ is much greater than one, rendering it optically thick.
The thickness of the layer$_{\tau = 1}$ in the filament ($d_{los, \tau=1}$) equals $d_{1}$, allowing us to observe a shallow region (see the green part in Figure\,\ref{fig:cartoon}(c)) beneath the upper layer of the filament with a filament temperature of $T_{1}$ ($\ll T_{0}$).
This observation corresponds to the brightness temperature at $t_{1}$ in Figure\,\ref{fig:cartoon}(a).
At $t_{2}$ in Figure\,\ref{fig:cartoon}(d), the filament continues to expand with the obscuration area of A$_{2}$ (> A$_{1}$).
The $\tau$ is still much greater than one, although the filament density decreases.
However, $d_{los, \tau=1}$ increases to $d_{2}$ due to the relationship $d_{los, \tau=1} \sim n_{e}^{-2}$.
Consequently, we can observe a deeper region within the filament with a brightness temperature of $T_{2}$ ($< T_{1}$), which corresponds to the brightness temperature at $t_{2}$ in Figure\,\ref{fig:cartoon}(a).
At $t_{3}$ in Figure\,\ref{fig:cartoon}(e), the filament continues to expand with $\tau$ still much greater than one, and its density continues to decrease.
Meanwhile, $d_{los, \tau=1}$ increases to $d_{3}$, corresponding to the center of the filament with the lowest temperature $T_{3}$.
The obscuration area (A$_{3}$) by the filament may have reached its maximum value at this time.
These may contribute to the maximum dimming depth observed at $t_{3}$ in Figure\,\ref{fig:cartoon}(a).
As time goes to $t_{4}$, at the end of or after the eruption, the filament density has significantly decreased due to extensive expansion, and $\tau$ drops below 1.0 again.
At this stage, we can observe a position below the filament with coronal emission, corresponding to $t_{4}$ in Figure\,\ref{fig:cartoon}(a).
Furthermore, the process illustrated from Figure\,\ref{fig:cartoon}(b) to \ref{fig:cartoon}(f) may also influence the variations of the brightness temperature as shown by the red symbols in Figures\,\ref{fig:Tvariation}(c) and \ref{fig:Tvariation}(f).
The variation of the optical depth from an optically thick regime to an optically thin one during a filament eruption is similar to the model results in the synthetic \lya\ waveband by \cite{2022A&A...665A..39Z}.

Filament eruptions are believed to occur frequently on other stars \citep[e.g.,][]{2024ApJ...961...23N}.
However, observational evidence for stellar filament eruptions and their associated CMEs remains limited \citep[e.g.,][]{2022ApJ...933...92C,2022A&A...663A.140L,2025ApJ...978L..32L}.
Conducting a radio dimming investigation with a sun-as-a-star analysis may offer a viable method for detecting stellar filament eruptions.

\section{Summary}
\label{sec:sum}

In the present paper, we present radio dimmings associated with two filament eruptions using the high-quality DART imaging observations in the meter and decimeter wavebands.
These dimmings manifest as significant emission depressions in the radio images in the frequency range of 150--450 MHz.
The observed radio dimmings are likely attributed to free-free absorption resulting from the obscuration of the solar corona by the eruptive filaments.
Additionally, in the AIA 304 \AA\ images, portions of the eruptive filaments appear as dark structures, indicating that they are predominantly composed of cool plasma.

The derived sun-as-a-star flux curves of the normalized brightness temperature across all frequencies also exhibit clear radio dimmings.
The dimming depths range from 1.5\% to 8\% of the pre-eruption level and show a negative correlation with frequencies and a positive correlation with filament areas.
In these two events, radio dimming becomes detectable in the sun-as-a-star flux curves when the eruptive filaments occupy more than 2\% of the solar disk area.
Our findings suggest that radio dimmings could serve as a new method for detecting stellar filament eruptions.
However, whether this method can distinguish between successful and failed filament eruptions requires further investigation.

\begin{acknowledgements}
This work was supported by the National Key R\&D Program of China No. 2021YFA0718600, NSFC grants 12425301, 12250006, 12303057, and China Postdoctoral Science Foundation No. 2021M700246.
H.C.C. was supported by NSFC grant 12103005.
M.M. acknowledges DFG grant WI 3211/8-1 and 3211/8-2, project number 452856778, and was supported by the Brain Pool program funded by the Ministry of Science and ICT through the National Research Foundation of Korea (RS-2024-00408396).
H.T. also acknowledges support from the New Cornerstone Science Foundation through the Xplorer Prize.
The authors thank Peijin Zhang for useful discussions.
AIA is an instrument onboard the Solar Dynamics Observatory, a mission for NASA's Living With a Star program.
CHASE mission is supported by the China National Space Administration.
ASO-S mission is supported by the Strategic Priority Research Program on Space Science, the Chinese Academy of Sciences, Grant No. XDA15320000.
We thank DART, SDO/AIA, CHASE, ASO-S, and SOHO/LASCO for providing data.
\end{acknowledgements}

\bibliographystyle{aa}
\bibliography{bibliography}

\end{document}